\documentstyle[12pt]{article}
\begin{document}
%======================================%
%<<<<<<<<<<<  DEFINITION  >>>>>>>>>>>>>%
%======================================%
\newcommand{\gsim}{\mbox{\raisebox{-1.0ex}{$\stackrel{\textstyle >}
{\textstyle \sim}$ }}}
\newcommand{\lsim}{\mbox{\raisebox{-1.0ex}{$\stackrel{\textstyle <}
{\textstyle \sim}$ }}}
\newcommand{\bfx}{{\bf x}}
\newcommand{\bfy}{{\bf y}}
\newcommand{\bfr}{{\bf r}}
\newcommand{\bfk}{{\bf k}}
\newcommand{\bkp}{{\bf k'}}
\newcommand{\order}{{\cal O}}
\newcommand{\beq}{\begin{equation}}
\newcommand{\eeq}{\end{equation}}
\newcommand{\beqn}{\begin{eqnarray}}
\newcommand{\eeqn}{\end{eqnarray}}
\newcommand{\lmk}{\left(}
\newcommand{\rmk}{\right)}
\newcommand{\lkk}{\left[}
\newcommand{\rkk}{\right]}
\newcommand{\lnk}{\left\{}
\newcommand{\rnk}{\right\}}
\newcommand{\call}{{\cal L}}
\newcommand{\calh}{{\cal H}}
\newcommand{\ppp}{\partial}
\newcommand{\xbar}{{\overline x}}
\thispagestyle{empty}
%======================================%
%<<<<<<<<<<<< TITLE PAGE >>>>>>>>>>>>>>%
%======================================%
\thispagestyle{empty}
{\baselineskip0pt
\rightline{\large\baselineskip16pt\rm\vbox to20pt{\hbox{UTAP-234}           
               \hbox{OCHA-PP-80}\hbox{RESCEU-21/96} 
               \hbox{June 1996}
\vss}}%
}
\vskip15mm
\begin{center}
{\large\bf Quantum Subcritical Bubbles}
\end{center}
\begin{center}
{\large Tomoko Uesugi and Masahiro Morikawa} \\
\sl{Department of Physics, Ochanomizu University, Tokyo 112, Japan}
\end{center}
\begin{center}
{\large Tetsuya Shiromizu} \\
\sl{Department of Physics, The University of Tokyo, Tokyo 113, Japan \\
and \\
Research Center for the Early Universe(RESCEU), \\ The University of Tokyo, 
Tokyo 113, Japan
}
\end{center}
%\begin{center}
%{\it }
%\end{center}
%======================================%
%<<<<<<<<<<<<< ABSTRACT >>>>>>>>>>>>>>>% 
%======================================%
\begin{abstract} 
We quantize subcritical bubbles which are formed in the weakly 
first order phase transition. We find that the typical size 
of the thermal fluctuation reduces in 
the quantum-statistical physics. 
We estimate the typical size and the amplitude of thermal fluctuations 
near the critical 
temperature in the electroweak phase transition using quantum 
statistical average. Furthermore based on our study, we give implication on 
the dynamics 
of phase transition. 
 
\end{abstract}
%\vfill
\vskip1cm

%\multicols{2}

%======================================%
%<<<<<<<<<<<< SECTION I  >>>>>>>>>>>>>>%
%======================================%
%\baselineskip25pt
\section{Introduction}
The scenario of baryogenesis in the electroweak scale\cite{ctm} is 
attractive because we can get the data of this energy scale from 
recent experiments of elementary particle physics. However, when one 
trys to explain the present baryon to entropy ratio in the minimal 
standard model, some serious problems exist. 
1)In order to generate sufficient baryon number, we need a deviation from 
thermal 
equilibrium. This can be solved if the phase transition is strongly
 first order with supercooling. 
However, in the minimal standard model,
electroweak phase transition is estimated to be weakly first order in the 
1-loop
approximation \cite{ew}.
2) CP-violation source from CKM phase is too small to explain present
baryon to entropy ratio.
3) We could not avoid washing out of baryon number by the sphaleron 
transition after the phase transition in the minimal standard model. 
Improvements and more precise analysis are now in progress to solve the 
above problems.  
In this paper, we devote to the analysis of the phase transition. 
To understand the dynamics of the electroweak phase 
transition, one may need to consider a behavior of the thermal fluctuation of
the scalar field. We study on a thermal fluctuation  using
subcritical bubbles which was first suggested by Gleiser et al\cite{gl}. 
In these analysis bubbles are treated as the classical object and 
assumed that the shape is spherical. 

Recently one of the present authors insisted that the fluctuation of the 
subcritical bubble 
should be treated quantum mechanically. 
So one must quantize the bubble and consider up to higher energy levels. 
In this situation, one should take into account the modification from the 
$O(3)$-symmetry of the bubble because the energy difference between each level 
is almost the
same as differential energy. Another reason that we need to take into account 
the above modification exists. In our previous 
studies we have convinced that 
the phase mixing is sufficiently attained in the experimentally 
allowed region on the minimal standard model. From the above reasons, the 
modification cannot be neglected. 

This paper is organized as follows. In Sec. II we summarize our previous 
studies. In Sec. III we quantize the collective coordinate 
of the subcritical bubble including the deviation form $O(3)$-symmetry and 
estimate the quantum-statistical averaged radius of the bubble. In Sec. IV 
we calculate the field amplitude inside the bubble. In the final section, we 
summarize our study. 

%======================================%
%              SECTION
%======================================%
\section{The  typical size of classical subcritical bubbles}
We summarize in this section our previous results on the classical bubbles.  
In the weakly first order phase transition, the time development of the
phase transition strongly depends on the thermal fluctuation.  
In the previous study \cite{shiromizu}, we have assumed 
%============< EQUATION >==============%
%
\begin{equation}
\phi (\bfx) = \phi_+ {\rm exp} \lkk -\frac{r^2}{R^2(t)} \rkk
\end{equation}
%
%======================================%
for the spatial profile of the thermal fluctuation. $\phi_+$ is the 
field value at the asymmetric vacuum phase. The Hamiltonian for the radius 
R(t) 
becomes  
%============< EQUATION >==============%
%
\begin{equation}
H(P,R)=\frac{P^2}{2M(R)}+\frac{2}{5}M(R),
\end{equation}
%
%======================================%
where 
%============< EQUATION >==============%
%
\begin{equation}
M(R)=\frac{15{\pi}^{3/2}{\phi}_+^2R}{8{\sqrt {2}}}.  
\end{equation}
%
%======================================%
This profile will be justified near the critical temperature at 
which two vacua degenerate  because 
almost all bubbles have thick walls. 
Here we neglected the term from the potential of the scalar 
field because that term vanishes 
near the critical temperature. Thus we obtained the typical size of 
thermal fluctuation by thermally averaging the radius;
%============< EQUATION >==============%
%
\begin{equation}
\langle R \rangle_T = \frac{\int dRdP R {\rm exp}[-\beta H(P,R) ]}
{\int dRdP {\rm exp}[-\beta H(P,R)]} \simeq \frac{2{\sqrt {2}}T}{\pi^{3/2}.  
\phi_+^2}
\end{equation}
%
%======================================%
Note that this is smaller than the correlation length. 
On the other hand, Gleiser et al. have assumed that the size of the bubble is 
equal to the correlation length\cite{gl}. We think that the typical 
size should be determined from the dynamics of the phase transition. 
However, we encounter with a serious problem: Can one treat the 
thermal fluctuation as the purely classical object? 
Actually, the number of state inside the bubble is too small;
%============< EQUATION >==============%
%
\begin{equation}
n(r \leq \langle R \rangle_T) \sim {\cal O}(1)
\end{equation}
%
%======================================%
Furthermore the condition for the decoherence, another measure for the system 
to be classical, is given by  
%============< EQUATION >==============%
%
\begin{equation}
\langle R \rangle_T \leq R_c:=\lmk \frac{27{\pi}^{3/2}}{{\sqrt {2}}T\phi_+^2} 
\rmk^{1/3}
\sim 0.084 {\rm GeV}^{-1} .
\end{equation}
%
%======================================%
On the other hand, $\langle R \rangle_T \sim 0.012{\rm GeV}^{-1}$, 
clearly, $ \langle R \rangle_T < R_c $. Thus, one cannot expect that 
the subcritical bubble is a classical object. 

From the above consideration one of the authors proposed that the fluctuation 
should be 
treated as quantum one with statistical fluctuation-dissipation in 
weakly first order phase transitions\cite{izumi}. 
In this paper, we first treat the subcritical bubble as purely quantum and 
then we statistically average the quantum bubbles. 
If we obtained, after our present analysis, no drastic deviation from the 
previous estimate, we 
would conclude again, without solving the Master equation for the 
density matrix, that 
the phase transition cannot accompany the supercooling.   
%======================================%
%<<<<<<<<<<   SECTION  >>>>>>>>>>>>>>>>%
%======================================%
\section{The typical size of quantum subcritical bubbles}
\subsection{Subcritical bubble with modification}

We notice that the energy difference is the same order of magnitude  
as the modification energy from the spherical symmetry 
in the statistical averaging. Hence, in advance, we 
consider the contribution of the modification from the spherical 
configuration. We express the spatial profile of the thermal
fluctuation as 
%============< EQUATION >==============%
%
\begin{equation}
\phi (\bfx) = \sum_{\ell , m} \phi_{\ell, m}(r) Y_{\ell, m}(\Omega),
\end{equation}
%
%======================================%
where $Y_{\ell, m}(\Omega)$ is the spherical harmonics. 
Further, we assume the thick wall
%============< EQUATION >==============%
%
\begin{equation}
\phi_{\ell, m}(r) = \phi_+ {\rm exp} \lkk -\frac{r^2}{R^2(t)} \rkk 
\end{equation}
%
%======================================%
for each components. Here $\phi_+$ denotes the vacuum expectation value 
of the scalar field in the broken phase.
The Hamiltonian becomes 
%============< EQUATION >==============%
%
\begin{equation}
 H(P,R)= H_0+\sum_{\ell =1}^\infty (H_{\ell}-H_0),
\end{equation}
%
%======================================%
where
%============< EQUATION >==============%
%
\begin{equation}
H_{\ell}(P,R)=\frac{P^2}{2M(R)}+\frac{2}{5}M(R)+\frac{16}{15} \ell 
(\ell +1) M(R) .
\end{equation}
%
%======================================%

\subsection{The quantization and the measure problem}

Let us quantize the above system. Expressing the arbitrariness in the operator 
ordering by the parameter b, we write the quantum Hamiltonian  
as follows;
%============< EQUATION >==============%
%
\begin{equation}
{\hat H}_{\ell}=-\frac{1}{aR}\frac{\ppp^2}{\ppp R^2}+
\frac{1}{bR^2}\frac{\ppp}{\ppp R} +c_{\ell}R,
\end{equation}
%
%======================================%
where $a:=2M(R)/R$ and $c_\ell :=[(2/5)+(16/15) \ell(\ell+1)]
(M/R)$. 

In the following discussion, 
we set $b/a=N$ and we concentrate on only two cases $N=1$ and $N=2$. 
Here one should be careful on the measure in the latter case.
The former case $N=1$ corresponds to 
the operator ordering;
%============< EQUATION >==============%
%
\begin{equation}
{\hat H}_{\ell}({\hat P},R)={\hat P}\frac{1}{2M(R)}{\hat P} +c_{\ell}R,
\end{equation}
%
%======================================%
where ${\hat P}:=-i \ppp_R$. Thus, the integral measure of this case 
is constant. 
The latter case $N=2$ corresponds to 
%============< EQUATION >==============%
%
\begin{equation}
{\hat H}_{\ell}({\hat z}_p,z)=-\frac{1}{2}{\hat z}^2_p +\cdot \cdot 
\cdot,
\end{equation}
%
%======================================%
where $z:=(R/R_0)^{3/2}$, $R_0:=(R/M)^{1/2}$ and ${\hat z}_p:=-i \ppp_z$. 
Hence the integral measure is
%============< EQUATION >==============%
%
\begin{equation}
\int^\infty_0 dz \propto \int^\infty_0 dR R^{1/2}.
\end{equation}
%
%======================================%

%======================================%
\subsection{The wave function}
%======================================%

Let us construct the wave function of the subcritical bubble.
The Schr\"odinger equation is given by 
%============< EQUATION >==============%
%
\begin{equation}
i\frac{\ppp}{\ppp t} \Psi_{\ell} (R,t) = {\hat H}_{\ell} 
\Psi_{\ell} (R,t). 
\end{equation}
%
%======================================%
For the stationary state ($ \Psi_\ell = \psi_\ell e^{-i E_\ell t} $, 
 ${\hat H}_\ell \psi_\ell = E_\ell \psi_\ell$ ),
the Schr\"odinger equation becomes
%============< EQUATION >==============%
%
\begin{equation}
-X_\ell ''+ \lkk \frac{1}{2N} (\frac{1}{2N} +1) x^{-2} +d_\ell \lmk 
x-\frac{\epsilon_\ell}{2d_\ell} \rmk^2 \rkk X_\ell = 
\frac{\epsilon_\ell^2}{4d_\ell}X_\ell , \label{eq:wave1}
\end{equation}
%
%======================================%
where we have defined
%============< EQUATION >==============%
%
\begin{equation}
X_\ell:= R^{-\frac{1}{2N}} \psi_\ell(R) ~~~~~{\rm and}~~~~~~x:=a^{1/2}R
\end{equation}
%
%======================================%
and 
%============< EQUATION >==============%
%
\begin{equation}
d_\ell:=\frac{c_\ell}{a}=\frac{1}{5}+\frac{8}{15}\ell ( \ell +1), ~~~~
\epsilon_\ell = \frac{E_\ell}{a^{1/2}} = \lmk \frac{R}{2M} \rmk^{1/2}
E_\ell
\end{equation}
%
%======================================%
In order to solve this equation, we take the approximation, 
%============< EQUATION >==============%
\beq
    x \gg \frac{\epsilon_\ell}{2d_\ell} \label{eq:kinji1}
\eeq
%======================================%
and we neglect the first term of the potential. Then the Schr\"odinger equation 
reduces to
%============< EQUATION >==============%
%
\begin{equation}
-X_\ell''+d_\ell \lmk x-\frac{\epsilon_\ell}{2d_\ell} \rmk^2 X_\ell =  
\frac{\epsilon_\ell^2}{4d_\ell}X_\ell. 
\end{equation}
%
%======================================%
This approximation will be examined in the later discussion. 

Now, we define 
%============< EQUATION >==============%
%
\begin{equation}
\xi:= d_\ell^{1/4} \lmk x- \frac{\epsilon_\ell}{2d_\ell} \rmk 
\end{equation}
%
%======================================%
and then eq. (20) becomes  
%============< EQUATION >==============%
%
\begin{equation}
-\frac{d^2X_\ell}{d \xi^2} +\xi^2 X_\ell =\frac{\epsilon_\ell^2}
{4d_\ell^{3/2}}X_\ell .
\end{equation}
%
%======================================%
As this equation is the same as that for the harmonic oscillator, 
the eigenfunctions and eigenvalues of this equation are given by 
%============< EQUATION >==============%
%
\begin{equation}
X_{\ell, n}(\xi) \propto (-1)^ne^{\frac{\xi^2}{2}} 
\frac{d^n e^{-\xi^2}}{d \xi^n}, \label{eq:nami1} 
\end{equation}
%
%======================================%
and
%============< EQUATION >==============%
%
\begin{equation}
\frac{\epsilon_{\ell, n}^2}{4d_\ell^{3/2}} =2n + 1,
\end{equation}
%
%======================================%
respectively. 

Using the above eigenvalues, we examine the approximation used on 
the potential under the assumption of (\ref{eq:kinji1}).
The original potential before the approximation is
%============< EQUATION >==============%
%
\begin{equation}
V_\ell(x):=\frac{1}{2N}\lmk \frac{1}{2N}+1 \rmk x^{-2}+d_\ell 
(x-x_0)^2,
\end{equation}
%
%======================================%
where $x_0:=\epsilon_{\ell , n}/2d_\ell$. Let us call the minimum  
value at which the first term coincides with the second term to be 
$x_*$. 

In the case of $N=1$ 
%============< EQUATION >==============%
%
\begin{equation}
x_* = \frac{x_0}{2} \lkk 1 + {\sqrt {1+\frac{2 \cdot 3^{1/2}}{2n+1}}} 
\rkk~~~~{\rm for}~~n=0, 1.
\end{equation}
%
%======================================%
This is larger than $x_0$ and the approximation brakes down. 
On the other hand, for $n \geq 2$,
%============< EQUATION >==============%
%
\begin{equation}
x_* = \frac{x_0}{2} \lkk 1 - {\sqrt {1 - \frac{2 \cdot 3^{1/2}}{2n+1}}} 
\rkk
\end{equation}
%
%======================================%
is smaller than $x_0$ and it means that the approximation is reasonable. 
In the case of $N=2$, in a similar way,  we can show 
that the approximation is reasonable only for $n \geq 1$. 

Thus, for low energy states, we have to use different approximation. 
The simplest remedy for that is as follows.  We first expand the potential
%============< EQUATION >==============%
%
\begin{equation}
V(x) \simeq V( \xbar ) + \frac{1}{2}V''(\xbar)(x-\xbar)^2+ \cdot 
\cdot \cdot,
\end{equation}
%
%======================================%
where $\xbar$ is determined by $V'(\xbar)=0$. Defining 
%============< EQUATION >==============%
%
\begin{equation}
\eta := \lkk \frac{1}{2}V''(\xbar) \rkk^{1/4}(x - \xbar),
\end{equation}
%
%======================================%
the Schr\"odinger equation for the stationary state becomes
%============< EQUATION >==============%
%
\begin{equation}
-\frac{d^2 X_\ell }{d \eta^2}+\eta^2 X_\ell 
=\frac{1}{\lkk \frac{1}{2}V''(\xbar) \rkk^{1/2}} 
\lkk \frac{\epsilon_\ell^2}{4 d_\ell }-V(\xbar) \rkk X_\ell.
\end{equation}
%
%======================================%
The solution is given by 
%============< EQUATION >==============%
%
\begin{equation}
X_{\ell, n}(\eta) \propto (-1)^ne^{\frac{\eta^2}{2}}\frac{d^n e^{-\eta^2}}
{d\eta^n} \label{eq:nami2}
\end{equation}
%
%======================================%
and
%============< EQUATION >==============%
%
\begin{equation}
\frac{1}{\lkk \frac{1}{2}V''(\xbar) \rkk^{1/2}} 
\lkk \frac{\epsilon_\ell^2}{4 d_\ell }-V(\xbar) \rkk =2n+1.
\end{equation}
%
%======================================%

Under the above two approximation, the wave function badly behaves at $R \sim 
0$. This violates the hermiticity of the 
Hamiltonian. However, this bad behavior simply comes from our approximation and 
should not be taken seriously. 
Actually we can easily see the true behavior near $R \sim 0$. At $x \ll 1$, 
the eq. (\ref{eq:wave1}) becomes 
%============< EQUATION >==============%
%
\begin{equation}
-\frac{d^2X_\ell}{dx^2}+\frac{1}{2N} \lmk \frac{1}{2N}+1 \rmk x^{-2}X_\ell
\approx  0.
\end{equation}
%
%======================================%
Assuming $X_\ell \propto x^n$, this becomes
%============< EQUATION >==============%
%
\begin{equation}
\lmk n-1-\frac{1}{2N} \rmk \lmk n+\frac{1}{2N} \rmk =0.
\end{equation}
%
%======================================%
Thus, the regular solution is $X_\ell \propto x^{1+1/2N} \propto 
R^{1+1/2N}$. The original wave function is given by 
$\psi_\ell (R) \propto R^{1+1/N}$. The first derivative is 
$\psi'(R) \propto R^{1/N}$. Therefore  actually the Hamiltonian 
is a hermitian operator. 

\subsection{The typical size} 

Now, we can calculate the typical size by the quantum-statistical 
averaging;
%============< EQUATION >==============%
%
\beq
 \langle R \rangle_Q = \frac{\sum_{\ell, n} e^{-\beta E_{\ell, n}}
      \langle R \rangle_{\ell, n}^Q }{\sum_{\ell, n} e^{-\beta E_{\ell, n}}}
\eeq
%
%======================================%
where
%============< EQUATION >==============%
%
\begin{equation}
\langle R \rangle_{\ell, n}^Q = \frac{\int^\infty_0dRR^{\alpha}R
|\psi_{\ell, n}(R)|^2 }{\int^\infty_0dRR^{\alpha}|\psi_{\ell, n}(R)|^2}.
\end{equation}
%
%======================================%
$R^\alpha$ being the integral measure; for $N=1$, $\alpha=0$, 
and for $N=2$, $\alpha=1/2$. The typical size of subcritical bubbles  
does depend on the integral measure, however after our approximation 
of the potential (\ref{eq:kinji1}), the dependence  on $N$ disappears.
Then, the effect of the integral measure only appears in the wave function with 
small $n$.  
Fortunately even for these wave functions, we can easily show that this effect 
to the typical size is small enough.  
We now estimate the typical size of a bubble using the effective potential in a 
finite temperature 
with the 1-loop approximation \cite{ew}. 
We use $m_W=80.6 {\rm GeV}$, $m_Z=91.2 {\rm GeV}$ 
and $m_t=174 {\rm GeV}$ for the $W$-boson, the $Z$-boson and the 
top-quark masses.  
And we set the mass of Higgs; $m_H=65{\rm GeV}$ that is the lower 
limit from the experiment\cite{pdg}. 
Accordingly, $\phi_+=49.4 {\rm GeV}$ and the critical 
temperature: $T_c = 98.3 {\rm GeV}$. 

The result of our numerical calculation for the typical size is, 
%============< EQUATION >==============%
%
\beqn
      \langle R \rangle_Q \sim 0.021 {\rm GeV}^{-1} 
\eeqn
%
%======================================%
near the critical temperature.
This size is larger than that of thermal average: 
$\langle R \rangle_T \sim 0.012{\rm GeV}^{-1}$. 

To clarify the difference between classical and quantum regime, 
we calculate the distribution function($ W (R_c) $);
%============< EQUATION >==============%
%
\begin{eqnarray}
W (R_c) & = & \int^\infty_{- \infty}dP R_c^{\alpha} e^{-i R_\Delta P} 
               W(P, R_c) \nonumber  \\
        & = &  \int^\infty_{- \infty}dP R_c^{\alpha}
               \int^\infty_{- \infty}dR_\Delta 
               e^{-i R_\Delta P} \rho\lmk R_c+\frac{1}{2}R_\Delta, 
                 R_c-\frac{1}{2}R_\Delta  \rmk \nonumber \\ 
        & = & R_c^\alpha \rho(R_c, R_c) ,
\end{eqnarray}
%
%======================================%
where $\rho$ is the density matrix given by 
%============< EQUATION >==============%
%
\begin{equation}
\rho\lmk R_c+\frac{1}{2}R_\Delta, R_c-\frac{1}{2}R_\Delta  \rmk 
=\rho (R, R')={\rm Tr}_{\ell, n} \lkk  e^{-\beta E_{\ell, n}} 
\psi^*_{\ell, n}(R')\psi_{\ell, n}(R) \rkk . 
\end{equation}
%
%======================================%
Under the present approximations, the final form of the distribution function 
does not 
depend on the measure;
%============< EQUATION >==============%
%
\begin{equation}
W(R_c)=R_c^\alpha \rho (R_c, R_c)=R_c{\rm Tr}_{\ell, n} \lkk  e^{-\beta 
E_{\ell, n}}X_{\ell, n}^2(R) \rkk. 
\end{equation}
%
%======================================%
We depicted profile of quantum distribution function $W(R)$ (solid line) and 
the classical distribution function (broken line) in figure 1. 
The location of the peak of $W(R)$ is larger than that of classical 
distribution function. This is because the fluctuations in quantum distribution 
function comes from 
quantum fluctuation as well as the statistical fluctuations which is expressed 
in the classical distribution function.

%======================================%
%<<<<<<<<<<<<<  SECTION >>>>>>>>>>>>>>>%
%======================================%

\section{Fluctuation strength in fields} 

While, fixing  $R$ to the above averaged value, we can estimate 
the fluctuation in the field value $\phi$. 
Following the manner of previous studies \cite{shiromizu} \cite{gl}, 
we set
%============< EQUATION >==============%
%
\beqn
     \phi_{lm}=A(t) \exp\left[-\frac{r^2}{\langle R\rangle^2_Q} \right].
\eeqn
%
%======================================%
In this case the Schr\"odinger equation becomes 
%============< EQUATION >==============%
%
\beq
 \left(-\frac{1}{2m}\frac{\partial^2}{\partial A^2}
    +f_{\ell}A^2 \right) \psi_{\ell}=E_{\ell} \psi_{\ell},
\eeq    
%
%======================================%    
where
%============< EQUATION >==============%
%
\beq
  f_{\ell}=\frac{m}{\langle R\rangle_Q^2}\left[\frac{3}{4}+
     2 \ell (\ell+1) \right] \ ,
  \ m=\frac{\pi^{3/2}\langle R\rangle_Q^3}{\sqrt{2}}
\eeq
%
%======================================%
and the wave function becomes
%============< EQUATION >==============%
%
\beqn
 \psi_{\ell, n}(a) & \propto & (-1)^n e^{\frac{a^2}{2}}\frac{d^n 
e^{-a^2}}{da^n}.
\eeqn
%
%======================================%
where we define
%============< EQUATION >==============%
%
\beq
  a=\left(2mf_{\ell} \right)^{\frac{1}{4}}A.  
\eeq
%
%======================================%
The fluctuation of $\phi$ can be calculated naively by 
%============< EQUATION >==============%
%
\beq
    \sqrt{\langle \phi^2 \rangle} = \sqrt{\frac{\sum_{\ell, n} e^{-\beta 
    E_{\ell, n}}\langle \phi^2 \rangle_{\ell, n}}{\sum_{\ell, n} 
   e^{-\beta E_{\ell, n}}}},
\eeq
%
%======================================%
where
%============< EQUATION >==============%
%
\beq
   \langle \phi^2 \rangle_{\ell, n} =
        \frac{\int dA A^2 \psi_{\ell, n}(A)}{\int dA \psi_{\ell, n}(A)}.
\eeq
%
%======================================%
The result of our numerical calculation becomes
%============< EQUATION >==============%
%
\beq
    \sqrt{\langle \phi \rangle^2} \sim  30.4 {\rm GeV}.
     \label{eq:rfix1}
\eeq
%
%======================================%
At the reflection point, $\phi=\phi_* =10.5 
{\rm GeV}$. 
%$\phi_*^+ =38.9971 {\rm GeV}$.
The fluctuation (\ref{eq:rfix1}) well exceeds $\phi_*$ 
 near the symmetric vacuum. 
This fact means that the true and false phases are well mixed at the scale 
typical in the fluctuations above the critical temperature.  
Therefore the use of 1-loop approximation of the effective potential is no 
longer 
valid.

%======================================%
%<<<<<<<<<<<<<< SECTION >>>>>>>>>>>>>>>%
%======================================%
\section{Summary}
In this paper, we quantized the subcritical bubbles and
using quantum statistics, we calculated the typical size of subcritical
bubbles as well as the typical fluctuation of fields. 

As a result of our calculations, we found that the typical size of subcritical 
bubbles is larger than that calculated by the classical thermal average, 
and that the size of the typical fluctuation of the field similarly well 
exceeds 
the first reflection point in our case. Therefore we may conclude that the 
phase transition cannot accompany supercooling. 
On the other hand, we must confess that the above study is quite qualitative. 
Furthermore, we could not justify whether the electroweak phase 
transition is first order or not 
as long as we use the 1-loop approximation for the effective potential, 
not {\it effective action}. 
 
Finally, we must mention the assumptions used here.
First, we have assumed the Gaussian distribution for the spatial profile 
of the thermal fluctuation as in the previous studies. However there is no 
dynamical justification on this Gaussian distribution. Second, in 
order to consider the deviation from the spherical symmetric configuration, we 
assumed that 
each component does not depend on the parameter $( \ell, n)$. 
We hope that we can justify the above assumptions in our quantitative study in 
the near future.

\vskip1cm

\centerline{\bf Acknowledgment}
TS thanks H. Sato, K. Sato and J. Yokoyama for their discussions and comments. 
This work was inspired by the referee's comment of progress of theoretical 
physics.
%This work was partly supported by Grant-in-Aid 
%for Scientific Research Fellowship, No.\ 2925 (TS).

%======================================%
%<<<<<<<<<<<< REFERENCES >>>>>>>>>>>>>>%
%======================================%

{\Large  Figure Captions}

\begin{enumerate}
\item{Fig.1:} Quantum statistical distribution function as a function 
of the radius $R$ of the subcritical bubble (solid line). Classical 
distribution function is also shown (broken line). The unit of the 
horizontal axis is ${\rm GeV}^{-1}$. 
\end{enumerate}

%\endmulticols
\end{document}